\documentclass[aps,pra,preprintnumbers,twocolumn,amsmath,amssymb,showpacs]{revtex4}
\usepackage{graphicx}
\usepackage{amsbsy}
\usepackage{dcolumn}
\usepackage{bm}
\usepackage{hyperref}
\usepackage[latin1]{inputenc}
\newcommand{\rs}{{\bf r}}

\newcommand{\ls}{{\bf l}}
\newcommand{\vs}{{\bf v}}

\newcommand{\ssm}{{\bf s}}
\newcommand{\ns}{{\bf n}}

\newcommand{\rc}{{\bf R}}

\newcommand{\ds}{{\bf d}}
\newcommand{\ks}{{\bf k}}

\newcommand{\ps}{{\bf p}}
\newcommand{\pc}{{\bf P}}
\newcommand{\fc}{{\bf F}}

\newcommand{\ac}{{\bf A}}
\newcommand{\bc}{{\bf B}}

\newcommand{\qs}{{\bf q}}

\newcommand{\xs}{{\bf x}}

\newcommand{\lc}{{\bf L}}

\newcommand{\ssc}{{\bf S}}

\newcommand{\eps}{{\varepsilon}}
\begin{document}
\title{Quantal statistical phase factor accompanying inter-change of two identical particles}
\author{Boyan D. Obreshkov$^{1,2}$}
\affiliation{Department of Physics, University of Nevada, Reno,
Nevada 89557, USA} \affiliation{Institute for Nuclear Research and
Nuclear Energy, Bulgarian Academy of Sciences, Tsarigradsko
chaussee 72, Sofia 1784, Bulgaria}
\date{\today}
\begin{abstract}
It is shown that the effects of particle statistics entail
reduction in the number of orbital degrees-of-freedom in
non-relativistic 2-particle systems from 6 to 5. The effect of
redundancy in the description of orbital motion is found to be in
correspondence to the multiplicative phase factor $(-1)^{2s}$
which accompany two-particle interchange, where $s$ is the spin of
one particle.

\end{abstract}
\maketitle
\section{Introduction}
The Pauli exclusion principle is basic principle in physics, in
particular it is usually related to the explanation of the shell
structure of atoms, conductivity of metals, stability of matter,
description of the properties of white dwarfs, and other phenomena
that are of experimental and theoretical interest.

A paper by Berry and Robbins has shown that the Pauli exclusion
principle may originate due to non-trivial kinematics of the
electronic spins \cite{spin}. Later, it was found that this
construction is not unique \cite{spin2}, and alternative
constructions of the exclusion principle are possible. However, an
earlier paper by Berry and Robbins \cite{spin3} has considered a
model $M=0$ spin system in external magnetic field, and derived
the phase factor $(-1)^S$ multiplying the wave-function of the
model system, when the direction of the magnetic field $\bc$ is
reversed. It was shown that the configuration space of the spin
system is equivalent to the configuration space of two identical
particles as constructed in Ref.\cite{topo}. The derivation in
Ref. \cite{spin3} shows that connection between spin and
statistics can be derived, rather than postulated. In different
papers \cite{bose,bose2}, a unique connection between spin and
statistics for spin $S=0$ bosons was derived, based on the
identification of the symmetric points $(\rs_1,\rs_2)$ and
$(\rs_2,\rs_1)$ in the configuration space of the two particles
and exploiting the continuity of the boson wave-function. In a
related paper \cite{Sudarshan}, the connection between spin and
statistics is shown to follow if the description of the dynamics
involves explicitly anti-commuting Grassmann variables. On the
other hand it is known that gauge structure appears in simple
classical and quantum mechanical systems \cite{Wilczek}. It has
been shown that long range forces in di-atom systems are mediated
by monopole-like gauge fields \cite{Zygelman}. Further elaboration
by Jackiw \cite{Jackiw1,Jackiw2} has shown that symmetries of
dynamic systems can be affected in presence of monopole gauge
fields. It is therefore reasonable to look for a connection
between spin and statistics in simple systems with two constituent
particles.

In Ref. \cite{reduc}, three-dimensional variational equation of
motion for $N$ Coulombically interacting electrons was derived,
which is different from the conventional $3N$-dimensional
many-body Schr\"{o}dinger equation. Unless otherwise stated, we
use atomic units $(e=m_e=\hbar=1)$.

The non-relativistic variational Schr\"{o}dinger equation of
motion for one active electron in presence of identical spectator
particle, and in absence of external (nuclear) forces is given by
\cite{reduc}
\begin{equation}
\left(-\frac{1}{2} \nabla_{\rs_1}^2 +\frac{g}{r_{12}} -\lambda
\right) \psi(\rs_1 \sigma_1,\rs_2\sigma_2) =0, \label{spect}
\end{equation}
where $g=1/2$, $r_{12}=|\rs_1-\rs_2|$ is the relative distance
between the spectator electron placed at $\rs_2$ and the active
electron located near $\rs_1$, $\sigma_1$ and $\sigma_2$ are the
components of the electronic spins $s=1/2$ on an arbitrary but
fixed spatial $z-$axis and $\lambda$ is unknown Lagrange
multiplier. The variational two-body fermion amplitude is given by
\begin{equation}
\psi(\rs_1 \sigma_1 , \rs_2 \sigma_2)=\langle {\rm vac} |
\Psi(\rs_1 \sigma_1) \Psi(\rs_2 \sigma_2) | \Psi \rangle,
\label{amplit}
\end{equation}
where $\Psi(\rs \sigma)$ is an anti-commuting fermion field
operator, $|{\rm vac} \rangle$ is the vacuum state of
non-interacting fermions and $|\Psi \rangle$ is unknown
state-vector of the interacting system of two electrons. Since the
fermion field operators anti-commute, then wave-functions of
electronic states in Eq.(\ref{amplit}) are anti-symmetric, i.e.
\begin{equation}
\psi(\rs_1 \sigma_1, \rs_2 \sigma_2) = -\psi(\rs_2 \sigma_2, \rs_1
\sigma_1), \label{Pauli}
\end{equation}
The subsidiary condition of Eq.(\ref{Pauli}) provides the
description of the dynamics of the spectator electron, which
otherwise is undetermined. That is because, if $\psi(\rs \sigma,
\rs' \sigma')$ is a solution of Eq.(\ref{spect}), then the
re-definitions
\begin{equation}
\psi(\rs_1 \sigma_1, \rs_2 \sigma_2)  \rightarrow  \sum_{\lambda_1
\lambda_2} U^{\sigma_1 \sigma_2}_{\lambda_1 \lambda_2}(\rs_2)
\psi(\rs_1 \lambda_1, \rs_2 \lambda_2) , \label{gauge}
\end{equation}
are also solutions of Eq.(\ref{spect}), where $U^{\sigma_1
\sigma_2}_{\lambda_1 \lambda_2}(\rs_2)$ is an arbitrary
$16$-element spinor matrix, which can depend locally on the
position-vector of the spectator electron, i.e. the phase and the
amplitude of the wave-function are not fixed by the equation of
motion alone. The physical solutions of Eq.(\ref{spect}) are fixed
by the Pauli's exclusion principle of Eq.(\ref{Pauli}), which
fixes the amplitude, the phase and the inter-relation between the
components of the electronic spins of the two-electron
wave-function, which otherwise remain arbitrary and undetermined.
However, since the state-vector $|\Psi\rangle$ is defined up to a
global phase, the relation of Eq.(\ref{Pauli}) can be written in
more general way as
\begin{equation}
\psi(\rs_1 \sigma_1, \rs_2 \sigma_2) = e^{i \theta} \psi(\rs_2
\sigma_2, \rs_1 \sigma_1),
\end{equation}
i.e. the two wave-functions are identical up to a global
phase-factor, independently on the fact that the fermion field
operators anti-commute. In the particular case, when $\theta=\pi$,
Eq.(\ref{Pauli}) is reproduced, i.e. the particle statistics phase
does not have a direct physical meaning, unless the system
undergoes a cycle in configuration space, such that $\theta$
represents the phase difference between initial and final state
wave-functions.

In the more general case of $N$-Coulombically interacting
electrons in presence of external one-body potential $U(\rs)$
\cite{reduc}, Eq.(\ref{spect}) together with the subsidiary
condition of Eq.(\ref{Pauli}), generalizes to one-electron
equation of motion
\begin{equation}
\left(-\frac{1}{2} \nabla_{\rs}^2 + U(\rs) + g \sum_{k=1}^{N-1}
\frac{1}{|\rs-\rs_k|} -\lambda \right)\psi(\rs
\sigma,\{\rs_k\sigma_k\})=0, \label{Fermion}
\end{equation}
together with $N!-1$ symmetry constraints
\begin{equation}
\psi(1,2, \ldots N) =\eta_P \psi[P(1),P(2), \ldots P(N)]
\label{bound}
\end{equation}
for anti-symmetry of the fermion wave-function due to the
Fermi-Dirac statistics,  by $(i)=(\rs_i \sigma_i)$ we have denoted
the coordinates of the $i$-th electron. The quantal statistical
phase-factors $\eta_P=(-1)^P$ accompanying the inter-change of
fermions are $+1$ if the permutation of the coordinates involves
even number of transpositions and $-1$ otherwise, and $N!$ is the
number of elements in the symmetric group $S_N$. The subsidiary
conditions for anti-symmetry define the structure of the Hilbert
space of $N$-electron wave-functions. The Hamilton equation of
motion, together with the subsidiary conditions that the
wave-function has to satisfy, correspond to Dirac's formulation of
the constrained quantum dynamics \cite{Dirac}.
The constraint Hamiltonian approach, has not found realization and
application in solving non-relativistic problems, such as the
calculation of the energy levels of the light hydrogen, helium and
lithium-like atoms and ions. The purpose of this paper is compute
the properties of the hydrogen and helium iso-electronic
sequences, and compare these results to the experiment.

\subsection{Theoretical formulation}
The state of the interacting $N$-electron system can be obtained
by solving a set of one-particle constraint equations for equal
sharing of the total energy by the particles
\begin{equation}
\chi_a |\psi \rangle = 0, \quad a =1, \ldots N \label{2ndclass}
\end{equation}
where
\begin{equation}
\chi_a = \frac{1}{2} \ps^2_a +  v_s(\rs_a) - \lambda
\end{equation}
are the operators of the constraints, $\lambda$ is uniform
Lagrange multiplier, $v_s(\rs)$ is the potential energy of the
active electron in an external field $U(\rs)$ and including the
repulsive Coulombic field of the spectator electrons
\begin{equation}
v_s(\rs_a)= U(\rs_a)+ \frac{1}{2} \sum_{b \ne a} r^{-1}_{ab},
\end{equation}
and $r_{ab}=|\rs_a-\rs_b|$ denote the relative distances between
particles. In addition there are $N$ first-class constraints for
identity of the particle spins, which are
\begin{equation}
\ssm_a^2 |\psi \rangle = s(s+1) |\psi \rangle, \quad a=1,\ldots N
\end{equation}
where $s=1/2$ is the spin of one electron, but otherwise $s$ can
be regarded arbitrary. The particle spin operators satisfy
canonical commutation relations
\begin{equation}
[(\ssm_a)_i,(\ssm_b)_j]=i \delta_{ab} \eps_{ijk} (\ssm_a)_k.
\end{equation}
The constraints in Eq.(\ref{2ndclass}) are all second-class, since
\begin{equation}
[\chi_a,\chi_b]=\frac{\rs_{ab}}{r^3_{ab}} \cdot \pc_{ab} \ne 0,
\end{equation}
where $\pc_{ab}=\ps_a+\ps_b$ is the momentum of the center-of-mass
motion of the electrons in the $(a,b)$-th pair. The constraints
are asymptotically first-class, since they decay with
inter-particle distances as $r_{ab}^{-2}$. The second-class
constraint system can be viewed as a result of gauge-fixing in an
extended first-class constraint system with gauge invariance. We
consider $N=2$, but we also consider $N=1$, since in this
particular case the present approach reduces exactly to the
one-particle Schr\"{o}dinger equation.

\subsection{Motion of one free electron}
By neglecting temporary the spin constraint, the Schr\"{o}dinger
equation of motion in momentum representation is given by
\begin{equation}
(\ps^2 - 2 E) \psi_E(\ps) = 0, \label{simpl}
\end{equation}
where the momentum $\ps$ of the particle is a multiplication
operator. The Schr\"{o}dinger equation is invariant under local
change of the phase of the wave-function
\begin{equation}
\psi(\ps) \rightarrow \psi(\ps) e^{i f (\ps)},
\end{equation}
since the momentum $\ps$ does not change
\begin{equation}
e^{i f (\ps)} \ps e^{-i f (\ps)}=\ps
\end{equation}
The phase of the free-particle wave-function is therefore
uncertain at each point in momentum space. Apart from the local
phase invariance, there is additional type of invariance of the
Schr\"{o}dinger equation under $s$-wave re-definition of the
wave-function, i.e.
\begin{equation}
\psi(\ps) \rightarrow \psi(\ps) + c(E) \delta(p-\sqrt{2 E})
\end{equation}
does not change the equation of motion. Therefore free-particle
states are defined up to an $s$-wave, which is a consequence of
the identity $(p^2-k^2) \delta(p-k) \equiv 0$. Therefore apart
from conventional symmetries of the one-particle Hamiltonian,
Schr\"{o}dinger equation exhibits two additional gauge symmetries,
it is invariant under local $U(1)$ phase transformations and under
$s$-wave redefinitions of the wave-function. The Hilbert space of
states is a quotient space,
\begin{equation}
{\cal H}_{{\rm phys}}= {\cal H} / {\cal H}_{{\rm s-wave}}.
\end{equation}
The states that are invariant under $s$-wave transformation of the
wave-function are
\begin{equation}
|\psi_{{\rm phys}} \rangle =|\psi\rangle -\langle E,l=0,m=0 | \psi
\rangle | E,l=0,m=0 \rangle,
\end{equation}
for some $|\psi \rangle \in {\cal H}$, i.e. $\langle \psi_{{\rm
phys}} | E,l=0,m=0 \rangle=0$, and hence the free-particle states
do not contain $s$-wave component. Two states are
gauge-equivalent, if they differ only by an $s$-wave
\begin{equation}
|\psi \rangle \sim |\psi \rangle + c(E) |E,l=0,m=0\rangle.
\end{equation}
To obtain free-particle wave-functions, gauge fixing-conditions $C
|\psi \rangle=0$ must be imposed, such that to select a
representative in each equivalence class.

We further can exploit the local phase-uncertainty of the
wave-function. The stationary group of the momentum $\ps$ is the
$SO(2) \cong U(1)$ group, i.e.
\begin{equation}
R_{\ps}(\chi) \ps R_{\ps}^{-1}(\chi) = \ps,
\end{equation}
where $R$ is a rotation operator and $\chi$ is the rotation angle
about the wave-vector $\ps$. This also reflects the operator
identity $\ps \cdot \ls=0$, where $\ls = \rs \times \ps$ is the
kinematic angular momentum. The eigenfunctions of the rotation
operator $-i
\partial_{\chi}$ are phase factors
\begin{equation}
\left.-i \partial_{\chi} e^{i \Lambda \chi}\right|_{\ps} = \Lambda
\left. e^{i \Lambda \chi}\right|_{\ps},
\end{equation}
where $\Lambda$ is some angular momentum quantum number and
$\chi=\chi(\ps)$. Each vector $\vs$ in the space $T_{\ps}$
tangential to the momentum  can be expanded
\begin{equation}
\vs = v_{\theta}(\ps) \partial_{\theta} + v_{\varphi}(\ps)
\partial_{\varphi},
\end{equation}
where the angles $(\theta,\varphi)$ specify the orientation of the
wave-vector $\ps$ in a space-fixed frame. The phase angle $\chi$
can be defined by the equation
\begin{equation}
\tan{\chi}= \left. \frac{v_{\theta}}{v_{\varphi}} \right|_{\ps},
\end{equation}
which locally changes in the interval $0 \le \chi \le 2 \pi$. In
view of arbitrariness of $\vs$, at each point $\ps$ of the
momentum space, we have the freedom to change locally the rotation
angle $\chi(\ps) \rightarrow \chi(\ps) + \alpha(\ps)$ without
affecting physical content. Since phase transformations of the
wave-function
\begin{equation}
\psi(\ps) \rightarrow \psi(\ps) e^{i \Lambda \chi(\ps)}
\end{equation}
do not change the equation of motion, and no gauge-fixing
conditions are imposed, the cyclic angle $\chi$ is a redundant
gauge degree-of-freedom and the momentum space of the particle
looks locally like $S^2 \times S^1$. The phase angle $\chi$,
however can affect the quantization of the orbital angular
momentum. The changes of the momentum under infinitesimal
variation $\ps \rightarrow \ps + d \ps$ induce Abelian $U(1)$
background gauge potential one-form $x = d \ps \cdot \xs(\ps)$
over the momentum space
\begin{equation}
d \ps \cdot \langle \Lambda(\ps) | i \nabla_{\ps} | \Lambda(\ps)
\rangle = d \ps \cdot \hat{\varphi} \frac{\Lambda}{p}
\cot{\theta},
\end{equation}
which maps the gradient of the phase of the wave-function, i.e.
measures the phase differences between wave-function values at
different points. When $\chi$ is re-defined locally
\begin{equation}
|\Lambda(\ps) \rangle \rightarrow |\Lambda(\ps) \rangle e^{-i
f(\ps)}, \quad \xs(\ps) \rightarrow \xs(\ps) + \nabla_{\ps} f(\ps)
\end{equation}
the induced displacement field $\xs(\ps)$ transforms as a gauge
field as it should, since the one-form $x(\vs) =\vs \cdot \xs$
takes values on vectors in $\vs \in T_{\ps}$, which are not
observable. If the momentum $\ps$ is displaced continuously along
a closed curve $C$, the phase-factor $\psi(\chi)=\exp{i \Lambda
\chi}$ satisfies the eigen-value equation $-i
\partial_{\chi} \psi(\chi)= \Lambda \psi(\chi)$ at each
point $\ps$. When the wave-vector returns to its original
direction, the wave-function is multiplied by a Berry's
phase-factor \cite{Berry}
\begin{equation}
\psi(\chi) \rightarrow \psi(\chi) \exp(i \oint_C x)=\psi(\chi)
\exp{i \gamma(C)},
\end{equation}
which generates a shift of the angle $\chi \rightarrow \chi+
\gamma(C)/\Lambda$, i.e. a gauge transformation. The gauge-field
exhibits Dirac string singularity along the entire $z$-axis, and
can not be defined globally. Singularity-free induced displacement
fields can be defined on two overlapping local patches as e.g.
\cite{Wilczek-book,monopole}
\begin{eqnarray}
& & \xs^N= \frac{\Lambda}{p} \frac{\cos{\theta}-1}{\sin{\theta}}
\hat{\varphi},
\quad R_N: 0 \le \theta < (\pi+\eps)/2 \nonumber \\
& & \xs^S = \frac{\Lambda}{p} \frac{\cos{\theta}+1}{ \sin{\theta}}
\hat{\varphi} \quad R_S: (\pi-\eps)/2 < \theta \le \pi
\end{eqnarray}
where the displacement field $\xs^N$ is regular on the north
hemi-sphere $R^N$, while $\xs^S$ is regular on the southern
hemi-sphere $R^S$. Near the equator $R^N \bigcap R^S$, these
fields are related by local gauge-transformation
\begin{equation}
\xs^S \rightarrow \xs^S -i e^{-2i \Lambda \varphi} \nabla_{\ps}
e^{2i \Lambda \varphi} = \xs^N .
\end{equation}
The induced displacement field is not rotation symmetric, since
under rotation $d \ps = \ns \times \ps$ it changes form.
Coordinate frame rotations are supplemented by local redefinitions
of the phase angle $\chi(\ps)$, such that rotation non-symmetric
terms be compensated, i.e. the following equation is satisfied
\begin{equation}
\ns \times \xs - \ns \times \ps \cdot \nabla_{\ps} \xs =
-\nabla_{\ps} f(\ps) \label{rot}
\end{equation}
where $f(\ps)$ is a compensating phase function. The gauge field
does not change form under Galilei boost transformations with
parameter $\delta \vs$, i.e.
\begin{equation}
G \xs (G^{-1} \ps) = \xs,
\end{equation}
for $G=I+ i \delta \vs \cdot \rs$. The gauge field leads to
rotation symmetric effects through its induced displacement field
strength two-form $F=dx=(\partial_i x_j(\ps) - \partial_j
x_i(\ps))dp^i \wedge dp^j$, its dual vector $F_i=\eps_{ijk}
F_{jk}$ or magnetic-like field is
\begin{equation}
\fc = \nabla_{\ps} \times \xs(\ps) = -\frac{\Lambda}{p^2}
\hat{\ps},
\end{equation}
Further, the kinematic angular momentum $\ls = \rs \times \ps$ of
the particle is supplemented by nonkinematic correction
\begin{equation}
\ls = -\ps \times \ds + \frac{\partial W}{\partial \ns},
\label{lcov}
\end{equation}
where $W=\Lambda \ns \cdot \hat{\ps}$ is the generator of local
phase transformations, corresponding to phase function
$f(\ps)=\Lambda \ns \cdot \hat{\ps}-\ns \times \ps \cdot \xs$ in
Eq.(\ref{rot}). The angular momentum $\Lambda \hat{\ps}$ is the
angular momentum stored in the gauge field. $U(1)$ gauge-invariant
Galilei boost generator $\ds$ is
\begin{equation}
\ds = \rs + \xs(\ps)
\end{equation}
and $\rs=i \nabla_{\ps}$ is the noninvariant canonical operator of
the position. The operator of the position $\ds$ is
non-commutative, and satisfies the relations
\begin{equation}
[d_i,d_j]=-i \Lambda \eps_{ijk} \frac{p_k}{p^3}, \quad [d_i,p_j]=i
\delta_{ij}
\end{equation}
that conserve canonical commutation relation between position and
momentum.  For spin-less particles $\Lambda=0$, the coordinates
commute. It is important however, that the Jacobi identity is not
satisfied
\begin{equation}
[[d_1,d_2],d_3]+ [[d_2,d_3],d_1]+[[d_3,d_1],d_2]=-4 \pi \Lambda
\delta^{(3)}(\ps)
\end{equation}
The gauge invariant representation of Galilei boost
transformations is based on the operator
\begin{equation}
G(\vs) = \exp(i \vs \cdot \ds),
\end{equation}
which generates transformation of the wave-function according to
\begin{equation}
G(\vs) \psi(\ps)= \exp(i \vs \cdot \ds) \exp(-i \vs \cdot \rs)
\psi(\ps-\vs).
\end{equation}
The product of the two exponentials is easily evaluated and given
by a line integral
\begin{equation}
\exp(i \vs_1 \cdot \ds) \exp(-i \vs_1 \cdot
\rs)=\exp\left(i\int_{\ps-\vs_1}^{\ps} d \qs \cdot \xs(\qs)
\right)
\end{equation}
connecting the frame $\ps-\vs_1$ to $\ps$. Application of second
Galilei transformation $G(\vs_2)$, shows that
\begin{equation}
G(\vs_1) G(\vs_2) = \exp\left[i \Pi(\ps;\vs_1,\vs_2 )\right]
G(\vs_1+\vs_2), \label{Galileo}
\end{equation}
where $\Pi(\ps;\vs_1,\vs_2 )=\oint_{\triangle} x=\Lambda
\Omega_{\triangle}$ is the Berry's phase \cite{Berry}, i.e. the
solid angle subtended by triangle $\triangle$ formed by the
vertices of the wave-vectors $\ps$, $\ps-\vs_1$ and
$\ps-\vs_1-\vs_2$, as seen from the rest frame $\ps={\bf 0}$ of
the particle. Therefore the composition law of the Galilei boost
transformations is in general non-associative, e.g.
\cite{Jackiw3}. The associativity of the Galilei boost
transformations is expressed by the equation
\begin{equation}
[G(\vs_1) G(\vs_2)] G(\vs_3)  = G(\vs_1) [G(\vs_2) G(\vs_3)].
\end{equation}
Taking into account Eq.(\ref{Galileo}), the associativity of
finite Galilei boost transformations is restored, if and only if
the flux $\oint_S F$ of the displacement field strength two-form
through a tetrahedron enclosing the wave-vector $\ps$ by the three
Galilei transformations, is quantizied according to $\Lambda=N/2$.
Therefore quantization of the helicity $\Lambda$ with half-integer
numbers is a consequence of associativity of finite Galilei boost
transformations. Canonical commutation relations between the
components of the $U(1)$ gauge-invariant rotation operator
$[l_i,l_j]=i \eps_{ijk} l_k$ hold for $\Lambda=N/2$. The square of
the angular momentum operator in Eq.(\ref{lcov}) can be written as
\begin{eqnarray}
& & \ls^2=-\frac{1}{\sin^2{\theta}} \left[ \sin{\theta}
\frac{\partial}{\partial \theta} \left( \sin{\theta}
\frac{\partial}{\partial \theta} \right) + \right. \nonumber
\\
& & + \left. \left( \frac{\partial}{\partial \varphi} +i \Lambda
(1-\cos{\theta}) \right)^2 \right] + \Lambda^2
\end{eqnarray}
and the rotation operator about the $z$-axis is
$l_z=-i\partial_{\varphi}+\Lambda$. Angular momentum
eigen-functions are determined by the equations
\begin{equation}
\ls^2 |l m \Lambda \rangle = l(l+1) |l m \Lambda \rangle, \quad
l_z | l m \Lambda \rangle = m | l m \Lambda \rangle,
\end{equation}
for $l=|\Lambda|,|\Lambda|+1,\ldots$ and $-l \le m \le l$.
Wave-functions given by sectional Wu-Yang monopole harmonics
\begin{equation}
Y_{lm\Lambda}(\theta,\varphi)=\langle \theta,\varphi | l m \Lambda
\rangle
\end{equation}
or more explicitly, these are given by means of Jacobi polynomials
$P^{(\alpha,\beta)}_n(z)$ as
\begin{eqnarray}
& & Y_{lm \Lambda}(\theta,\varphi)=N_{lm} (1-z)^{-(\Lambda+m)/2}
(1+z)^{-(\Lambda-m)/2} \times \nonumber
\\
& & \times P^{(-\Lambda-m,-\Lambda+m)}_{l+m}(z) e^{i (\Lambda+m)
\varphi},
\end{eqnarray}
where $z=\cos{\theta}$ and $N_{lm}$ are normalization constants.
Components of the gauge-invariant rotation operator $\ls$ satisfy
canonical commutation relations $[l_i,l_j]=i \eps_{ijk} l_k$, and
therefore are connected to the Wigner's rotation functions by
\begin{equation}
Y_{lm\Lambda}(\theta,\varphi) = D^l_{\Lambda
m}(-\varphi,\theta,\varphi)=\langle l \Lambda | e^{-i \varphi l_z}
e^{i \theta l_y} e^{i \varphi l_z} | l m \rangle.
\end{equation}
The sign of $\Lambda$, ${{\rm sign}}(\Lambda)=\pm 1$ distinguishes
left-handed from right-handed rotations, which commute. The
wave-function is $\chi$-independent, single particle states
labelled by four quantum numbers
\begin{equation}
|\psi \rangle = |Elm\Lambda \rangle
\end{equation}
and $E$ is the kinetic energy of the particle. The states with
$|\Lambda|=0,1,2\ldots$ form representation of the rotation group
of integer angular momentum. For $\Lambda=0$, they reduce to the
conventional spherical harmonics $Y_{lm}(\hat{\ps})$. The states
corresponding to half-integer angular momentum $|\Lambda|=1/2,
3/2, \ldots$ define spinor representations of the rotation group.
For instance, a doublet of wave-functions corresponding to
$l=\Lambda=1/2$ is given by means of half angles
\begin{equation}
Y_{\frac{1}{2} \frac{1}{2}
\frac{1}{2}}(\theta,\varphi)=-\sin{\frac{\theta}{2}} e^{i
\varphi}, \quad Y_{\frac{1}{2} -\frac{1}{2}
\frac{1}{2}}(\theta,\varphi) = \cos{\frac{\theta}{2}},
\end{equation}
on the northern hemi-sphere of the momentum space. Second doublet
of wave-functions with $\Lambda=1/2$ with support on the southern
hemi-sphere is obtained by spatial inversion $\theta \rightarrow
\pi-\theta, \varphi \rightarrow \varphi+\pi$. Second group of
left-moving helicity eigen-states of $\Lambda=-1/2$ is obtained by
complex conjugation of wave-functions of right-handed particle
states.

// (up to here)

\subsection{Motion of two free electrons.}
The Schr\"{o}dinger equation of motion for one free electron in
presence of identical spectator electron is,
\begin{equation}
\left(\ps_1^2  -\lambda \right) \psi(\rs_1 \sigma_1,\rs_2
\sigma_2) =0, \label{eom1}
\end{equation}
where $\ps_1=-i \nabla_{\rs_1}$ is the momentum of the active
electron. The interchange of the particles' position vectors and
spins $(\rs_1 \sigma_1) \leftrightarrow (\rs_2 \sigma_2)$ leads to
identical description of the motion of the spectator electron
\begin{equation}
\left(\ps_2^2  -\lambda \right) \psi(\rs_2 \sigma_2,\rs_1
\sigma_1) =0. \label{eom2}
\end{equation}
Since the interchange of particles changes only the sign of the
wave-function, the comparison of Eq.(\ref{eom1}) with
Eq.(\ref{eom2}) shows that the kinetic energies of the two
particles are equal, i.e.
\begin{equation}
\ps_1^2 |\psi\rangle = \ps_2^2 |\psi\rangle = 2 \lambda
|\psi\rangle  \label{kinet2}
\end{equation}
the particles move such that to conserve identical their
de-Brogile wave-lengths $\lambda_{{\rm dB}}= 2\pi /\sqrt{2
\lambda}$. In addition particles exhibit identical spins $s$, i.e.
\begin{equation}
\ssm_1^2 |\psi\rangle = \ssm_2^2 |\psi\rangle = s(s+1)
|\psi\rangle \label{spin2}.
\end{equation}
The effects of particle interchange do not involve exchanging
energy and momentum and can be represented by rotations of unit
vectors $\hat{\ps}_1$ and $\hat{\ps}_2$ along with rotations of
half-integer spins, since magnitudes of momenta $p_1=p_2=\sqrt{2
\lambda}$ and spins $s$ are not relevant for the description of
the effect of particle interchange, i.e. if for instance the
momenta are simultaneously scaled according to $\ps_1 \rightarrow
e^{\theta} \ps_1$ and $\ps_2 \rightarrow e^{\theta} \ps_2$, then
the constraint equation remains unchanged. The equation of motion
for the active electron is invariant under bi-local phase change
of the wave-function in momentum space
\begin{equation}
\psi(\ps_1,\ps_2) \rightarrow \psi(\ps_1,\ps_2)  e^{i
f(\ps_1,\ps_2)}
\end{equation}
and is invariant under $s$-wave transformation
\begin{equation}
\psi(\ps_1,\ps_2) \rightarrow \psi(\ps_1,\ps_2) + \delta(p_1-k)
c(\ps_2),
\end{equation}
where $c(\ps_2)$ is an arbitrary function of the spectator
momentum. For comparison, the conventional two-particle
Schr\"{o}dinger equation
\begin{equation}
(\ps_1^2 + \ps_2^2 -2 E) \psi(\ps_1,\ps_2)=0
\end{equation}
is invariant under bi-local phase transformation, and similar
$s$-wave transformation
\begin{equation}
\psi(\ps_1,\ps_2) \rightarrow \psi(\ps_1,\ps_2) + c
\delta(p_1-k)\delta(p_2-k).
\end{equation}
We further could separate the orbital from the spin variables, by
demanding that the total wave-function be an eigen-function of
total spin ${\bf S}=\ssm_1 +\ssm_2$, together with its projection
$M$ onto a space-fixed unit-vector $\hat{\pc}$, i.e.
\begin{equation}
{\bf S}^2 |\psi_S \rangle = S(S+1) |\psi_S \rangle, \quad
\hat{\pc} \cdot {\bf S} |\psi_S \rangle = M |\psi_S \rangle
\end{equation}
and the wave-function is
\begin{equation}
\psi(\rs_1 \sigma_1 , \rs_2 \sigma_2)= \psi_S(\rs_1,\rs_2)
C^{SM}_{s \sigma_1, s \sigma_2}.
\end{equation}
The Clebcsh-Gordan coefficient changes under interchange of spins
$\sigma_1 \leftrightarrow \sigma_2$ as
\begin{equation}
C^{SM}_{s \sigma_1 s \sigma_2}=(-1)^{2s-S} C^{SM}_{s \sigma_1,s
\sigma_2}.  \label{clebsch}
\end{equation}
and implies that under interchange of spatial coordinates
\begin{equation}
\psi_S(\rs_1,\rs_2)=(-1)^S \psi_S(\rs_2,\rs_1),
\end{equation}
the wave-function is multiplied by the phase-factor $(-1)^S$. We
further change the individual coordinates to collective
coordinates for the relative $\rs=\rs_1-\rs_2$ and center-of-mass
motion $\rc=(\rs_1+\rs_2)/2$. The momenta, which are conjugate to
these coordinates are $\ps=-i \nabla_{\rs}$ and $\pc=-i
\nabla_{\rc}$, respectively.  The equation of motion reads
\begin{equation}
\left(\frac{1}{2} \ps^2 + \frac{1}{2} \ps \cdot \pc + \frac{1}{8}
\pc^2  -\lambda \right) \psi_S(\rc,\rs) =0, \label{active}
\end{equation}
and the boundary condition of Eq.(\ref{Pauli}) now reads
\begin{equation}
\psi_S(\rc,\rs)=(-1)^S \psi_S(\rc,-\rs),  \label{boundary2}
\end{equation}
The inter-change of spatial coordinates $\rs \rightarrow -\rs$ in
Eq.(\ref{active}) leads to the equation of motion for the
spectator electron
\begin{equation}
\left(\frac{1}{2} \ps^2 - \frac{1}{2} \ps \cdot \pc + \frac{1}{8}
\pc^2  -\lambda \right) \psi_S(\rc,\rs) =0, \label{spectt}
\end{equation}
Since Eq.(\ref{active}) and Eq.(\ref{spectt}) are satisfied
simultaneously, we have single first-class constraint on the
dynamics
\begin{equation}
\pc \cdot \ps |\psi_S \rangle =0, \label{FD}
\end{equation}
that particles share the kinetic energy in equal way, and
therefore can not be distinguished. The presence of spectator
particle is non-trivial, since it constraints the wave-function of
the two-particle state. Hamiltonians in Eq.(\ref{active}) and
Eq.(\ref{spectt}) together with the constraint of Eq.(\ref{FD})
are translation and rotation invariant. We further constraint the
wave-function to be an eigen-function of the conserved momentum
$\pc$ of the center-of-mass motion
\begin{equation}
\psi_S(\rc,\rs)= e^{i \pc \cdot \rc} \psi_{S,\pc} (\rs),
\end{equation}
and re-write the equation for the relative motion as
\begin{equation}
\left(\ps^2 -k^2  \right) \psi_{S,\pc}(\rs) = 0,
\label{active-new}
\end{equation}
where $k^2=2 \lambda -P^2/4$. The relative wave-function is
subject to constraint for anti-symmetry
\begin{equation}
\psi_{S,\pc}(\rs)=(-1)^S \psi_{S,\pc}(-\rs).
\end{equation}
By neglecting effects of spin, the solution of Eq.(\ref{spectt})
can be written as
\begin{equation}
\psi_{\pc}(\rs)=e^{i \ks \cdot \rs} e^{i f_{\pc}(\rs)},
\end{equation}
i.e. the phase of the unconstrained Schr\"{o}dinger's
wave-function is re-defined locally by the constraint for equal
sharing of kinetic energy. Since the particles are free, the
phase-function $f_{\pc}(\rs)$ is a linear function of the relative
coordinate $\rs$, i.e. $f(\rs)= \ac_{\pc} \cdot \rs$, where
$\ac_{\pc}$ is a constant vector. The non-trivial solution for the
compensating vector is $\ac_{\pc} = -(\ks \cdot
\hat{\pc})\hat{\pc}$. The wave-function is therefore given by
\begin{equation}
\psi_{\pc}(\rs)= e^{i [\ks \cdot \rs - (\ks \cdot \hat{\pc})(\rs
\cdot \hat{\pc})]},  \label{psi}
\end{equation}
and the kinetic energy of relative motion is $\eps=\ks^2 - (\ks
\cdot \hat{\pc})^2$, indicating that the component of the relative
momentum $\ks$ on the direction of propagation $\hat{\pc}$ is
redundant. The kinematic constraint for equal sharing of kinetic
energy annihilates the wave-function
\begin{equation}
\delta_{\eps} \psi_{\pc} (\rs) = \eps \pc \cdot \ps \psi_{\pc}
(\rs) = 0
\end{equation}
where $\eps$ is an uniform gauge parameter, i.e. under the
transformation
\begin{equation}
\psi_{\pc}(\rs) \rightarrow \psi_{\pc}(\rs) + \delta_{\eps}
\psi_{\pc}(\rs)\approx \psi(\rs + \eps \pc)
\end{equation}
the two-electron wave-function remains unchanged. Furthermore the
uniform parameter $\eps$ can be "gauged" into a function
$\eps=\eps(\rs)$. The particle identity constraint is a generator
of canonical transformations of the variables in the dynamic
system, the relative coordinate is gauge-dependent and changes as
\begin{equation}
\delta_{\eps} \rs |\psi_{\pc} \rangle = -i \eps[\rs, \pc \cdot
\ps] |\psi_{\pc} \rangle =\eps \pc |\psi_{\pc} \rangle ,
\label{coord}
\end{equation}
i.e. acquires a longitudinal correction. The property of
sign-change of the vector $\rs=\rs_1-\rs_2$ inter-connecting the
particles under interchange is not unique, since the physically
equivalent position vector $\rs+ \eps \pc \rightarrow -\rs + \eps
\pc$ does not change sign, unless the transformation $\rs_1
\leftrightarrow \rs_2$ is supplemented by reversal of the sign of
$\pc$. However, the momentum of relative motion is unchanged, i.e.
$\delta_{\eps} \ps |\psi_{\pc} \rangle =0$, i.e. the interchange
of momenta $\ps \rightarrow -\ps$ is gauge-invariant
transformation. The relative angular momentum $\ls = \rs \times
\ps$ is gauge-dependent
\begin{equation}
\delta_{\eps} (\rs \times \ps) |\psi_{\pc} \rangle =  \eps \pc
\times \ps |\psi_{\pc} \rangle \ne 0,
\end{equation}
i.e. $\ls \sim \ls + \pc \times \ps$ are equivalent as operators
and can be identified, in the same way $\lc=\rc \times \pc \sim
\lc + \ps \times \pc$ can be identified. The total angular
momentum ${\bf J}=\lc+\ls+\ssc$ is gauge-invariant. If the
position vector interconnecting the particles is resolved as
\begin{equation}
\rs =\rs_{\bot} + \rs_{||},
\end{equation}
where $\rs_{||}= \hat{\pc} (\hat{\pc} \cdot \rs)$ is the
projection onto the propagation wave-vector, while $\rs_{\bot}$ is
the rejection, then the gauge invariance of the wave-function is
expressed by its independence on the projection $\rs_{||}$ and the
rejection coordinate $\rs_{\bot}$ is gauge-invariant coordinate.

To take into account more accurately the effect of spin $S$, the
operator of the constraint is resolved
\begin{equation}
\pc \cdot \ps = P (p_x \sin{\Theta} \cos{\Phi} + p_y \sin{\Theta}
\sin{\Phi} + p_z \cos{\Theta}) ,
\end{equation}
where $(P,\Theta,\Phi)$ are the spherical coordinates of the
propagation wave-vector. The inter-particle position vector is
further resolved in a local basis specified by the propagation
wave-vector $\hat{\pc}$ as
\begin{equation}
\rs = r_P \hat{e}_{\pc} + r_{\Theta} \hat{e}_{\Theta} + r_{\Phi}
\hat{e}_{\Phi}.
\end{equation}
The rotation matrix which gives the change of coordinates is
\begin{equation}
\left(
\begin{array}{c}
r_P \\
r_{\Theta} \\
r_{\Phi} \\
\end{array}
\right) = \left(
\begin{array}{ccc}
\sin{\Theta} \cos{\Phi} & \sin{\Theta} \sin{\Phi} & \cos{\Theta}
\\
\cos{\Theta} \cos{\Phi} & \cos{\Theta} \sin{\Phi} & -\sin{\Theta}
\\
-\sin{\Phi} & \cos{\Phi} & 0
\end{array}
\right) \left(
\begin{array}{c}
x \\
y \\
z
\end{array}
\right)
\end{equation}
Similarly the momentum of relative motion is expanded over this
basis
\begin{equation}
\ps = p_P \hat{e}_{\pc} + p_{\Theta} \hat{e}_{\Theta} + p_{\Phi}
\hat{e}_{\Phi}
\end{equation}
At each point $\pc$, cylindrical coordinates are introduced
\begin{equation}
\rho = \sqrt{r_{\Theta}^2 + r_{\Phi}^2 }, \quad
\tan{\varphi}=\frac{r_{\Phi}}{r_{\Theta}}, \quad z = r_{P}
\end{equation}
The subsidiary condition for equal sharing of kinetic energy takes
simple form
\begin{equation}
\partial_z \psi_{\pc,S}(\rho,\varphi,z)=0,  \label{z-constr}
\end{equation}
i.e. wave-function is independent on the longitudinal coordinate
$z$, and is an eigen-function of the operator of the momentum
$p_z=p_P$ with eigenvalue $k_z=k_P=0$. For each fixed propagation
wave-vector $\pc$, the wave-function satisfies the equation
\begin{equation}
\left(\partial^2_{\rho}  + \frac{1}{\rho} \partial_{\rho}  +
\frac{1}{\rho^2} \partial^2_{\varphi} +  k^2 \right)
\psi_{\pc,S}(\rho,\varphi)  = 0,  \label{cyl}
\end{equation}
and $E=k^2$ is the kinetic energy of relative motion. Under
$\pi$-re-definition of the fiber angle $\varphi$, the particles
interchange, and their wave changes according to
\begin{equation}
\psi_{\pc,S}(\rho,\varphi)=e^{i S \pi}
\psi_{\pc,S}(\rho,\varphi+\pi).   \label{bc-free-cyl}
\end{equation}
The solution of Eq.(\ref{cyl}) is separable $R(\rho)
\phi(\varphi)$ in cylindrical coordinates, and periodic Bloch-type
boundary condition in Eq.(\ref{bc-free-cyl}) fixes the solution as
\begin{equation}
\psi_{\pc,S}(\rho,\varphi,\sigma,\sigma') = J_{\Lambda}(k\rho)
e^{i \Lambda \varphi} C^{SM}_{s \sigma, s \sigma'}
\end{equation}
where $\Lambda=S {{\rm mod}}(2 \hbar)$ is the helicity, which is
analogue of a Bloch quasi-angular momentum, $J_{\Lambda}(z)$ are
the Bessel functions of integer order and $\pi$ is a
characteristic angular inter-change period. The sign of $\Lambda$
determines left or right helicity eigen-states. The inter-change
of particles $\rs \rightarrow -\rs$ can be expressed by
\begin{equation}
\psi_{\pc}(\varphi+\pi,\sigma', \sigma)= e^{i \pi S}
\psi_{\pc}(\varphi,\sigma',\sigma)= (-1)^{2s}
\psi_{\pc}(\varphi,\sigma,\sigma').  \label{spin-stat}
\end{equation}
where we have used the symmetry property of the Clebsch-Gordan
coefficient  in Eq.(\ref{clebsch}). Eq.(\ref{spin-stat}) is only a
consequence of the solution of the equations of motion, i.e. if we
project the state vector on exchanged configurations $\langle
\varphi+\pi,\sigma',\sigma| \psi \rangle=(-1)^{2s} \langle
\varphi, \sigma, \sigma' | \psi \rangle$, the multiplicative
particle-statistics phase-factor appears automatically. Therefore,
the particle inter-change is a gauge transformation of the fiber
angle $\varphi \rightarrow \varphi + \pi$, which partially
compensates the effect of rotation of the phase of the spin
wave-function. The "local"-type quantization of $\Lambda$ by Bloch
type boundary condition, is inappropriate, since the phases of
wave-functions evaluated at different points can not be compared,
which reflects the uncertainty of the phase of the momentum space
wave-function.

// (up to here)

The effect of particle identity is shown to entail reduction in
the number of the initial six orbital degrees of freedom to five.

\subsection{Two Coulombically interacting electrons in absence of external forces}
We consider the problem for the Coulomb scattering of the two
particles, when there is no source of external forces.
Furthermore, the spin constraints are represented very
approximately by Bloch-type boundary condition. Since the
Coulombic interaction $r^{-1}_{12}$ is invariant under
inter-change of particles, then nothing principal changes as
compared to the case of motion of free electrons. The Hamiltonian
of the active electron is
\begin{equation}
h_{\rs_1} = -\frac{1}{2} \nabla_{\rs_1}^2 + g r_{12}^{-1} ,
\label{hsing}
\end{equation}
By inter-changing the coordinates $\rs_1 \leftrightarrow \rs_2$,
we obtain the Hamiltonian for the motion of the spectator particle
\begin{equation}
h_{\rs_2} = -\frac{1}{2} \nabla_{\rs_2}^2 + g r_{21}^{-1},
\end{equation}
where $g=1/2$. Using Eq.(\ref{Pauli}) that the fermion
wave-function only changes sign upon inter-change of particles, we
obtain that
\begin{equation}
h_{\rs_1} \psi(\rs_1,\rs_2) = h_{\rs_2} \psi(\rs_1,\rs_2) =
\lambda \psi(\rs_1,\rs_2), \label{identical}
\end{equation}
i.e. the particles are precisely identical, since Hamiltonians
$h_{\rs_1}$ and $h_{\rs_2}$ exhibit common eigenvalue $\lambda$.
Therefore, equations of motion are consistent only if one-particle
Hamiltonians commute with each other, i.e. $[h(1),h(2)] |\psi
\rangle = 0$, which leads to a consistency condition
\begin{equation}
(\fc_{12} \cdot \ps_1 - \fc_{21} \cdot \ps_2 ) |\psi \rangle = 0,
\label{screening}
\end{equation}
where
\begin{equation}
\fc_{12} = \frac{\rs_{12}}{r_{12}^3} = - \fc_{21}
\end{equation}
is the repulsive Coulomb force of interaction between the two
particles. Eq.(\ref{screening}) shows that electrons move such
that to screen (compensate) the excess Coulombic force in the
direction of the total momentum $\pc=\ps_1+\ps_2$. This result has
simple classical analogue, since the above equation reads
\begin{equation}
\ps_1 \cdot \dot{\ps}_1 -\ps_2 \cdot \dot{\ps}_2 =0,
\end{equation}
and the difference of the kinetic energies of the two particles is
a constant of motion
\begin{equation}
\frac{d}{dt} (\ps_1^2 - \ps_2^2) = 0.
\end{equation}
When this difference is vanishing, the particles are precisely
identical, otherwise they can be distinguished trivially.
Furthermore, if the active electron at point $\rs_1$ changes its
momentum due to the Coulomb force of his partner positioned at
$\rs_2$, the spectator particle changes its momentum in strictly
proportional way, such that the excess force in the direction of
the motion of the center-of-mass $\pc$ is compensated. The
supplementary condition for screening in Eq.(\ref{screening}) of
the mutual excess Coulomb forces can be written more simply as
\begin{equation}
\rs_{12} \cdot (\nabla_1+\nabla_2) | \psi \rangle = 0.
\end{equation}
By means of Eq.(\ref{identical}), we also have that
$[h(1)-h(2)]|\psi \rangle=0$, which is a constraint for equal
sharing of kinetic energy by the particles, i.e.
\begin{equation}
[\ps_1^2 - \ps_2^2 ] |\psi \rangle =0.
\end{equation}
The constraint for screening of the repulsive inter-particle
Columbic force can be viewed as a gauge-fixing condition for the
invariance generated by the constraint of equilibration of the
de-Brogile wave-lengths $\ps_1^2=\ps_2^2$. Re-introducing the
collective coordinates for relative $\rs =\rs_1-\rs_2$ and
center-of-mass motion $\rc=(\rs_1+\rs_2)/2$, with the
corresponding momenta $\ps =-i \nabla_{\rs}$ and $\pc=-i
\nabla_{\rc}$. Further, the center-of-mass motion is uniform and
we impose explicitly three additional constraints for the
conservation of the momentum of center-of-mass $\pc=(P_x,P_y,P_z)$
motion
\begin{equation}
- i \nabla_{\rc} \langle \rc |\psi \rangle = \pc \langle \rc |\psi
\rangle,
\end{equation}
i.e. the two-electron state is an eigen-state characterized by the
momentum $\pc$, i.e.
\begin{equation}
\psi = e^{i \pc \cdot \rc} \psi_{\pc}(\rs).
\end{equation}
By separating the center-of-mass motion, the Hamiltonian for the
relative motion of the two particles becomes
\begin{equation}
H_{rel}= \ps^2 + |\rs|^{-1},
\end{equation}
subject to the supplementary condition for screening $\hat{\pc}
\cdot \rs |\psi_{\pc,S} \rangle =0$ and for equal sharing of
kinetic energy
\begin{equation}
\hat{\pc} \cdot \ps |\psi_{\pc,S} \rangle =0,  \label{kin}
\end{equation}
and $\hat{\pc}$ is a unit vector in the direction of propagation
of the center-of-mass motion. In Cartesian coordinates
$\rs=(r_P,r_{\theta},r_{\Phi})$ with $z$-axis parallel to the
propagation wave-vector, the pair of supplementary conditions
become
\begin{equation}
r_P \langle \rs |\psi_{\pc,S} \rangle =0 , \quad \frac{1}{i}
\frac{\partial}{\partial r_P} \langle \rs |\psi_{\pc,S} \rangle
=0,
\end{equation}
which is a pair of second-class constraints, that show that the
longitudinal relative coordinate $r_P$ is locally redundant. The
reduced Hamiltonian for the planar orbital motion of the internal
degrees-of-freedom simplifies as
\begin{equation}
H_{rel}= p_{\Phi}^2 + p_{\Theta}^2 + \frac{1}{\sqrt{ r_{\Theta}^2
+ r_{\Phi}^2 } },
\end{equation}
The wave-function is subject to the boundary condition
\begin{equation}
\psi_{S,\pc}(r_{\Theta},r_{\Phi})=(-1)^S
\psi_{S,\pc}(-r_{\Theta},-r_{\Phi}).
\end{equation}
Introducing the cylindrical coordinates $r_{\Theta}=\rho
\cos{\varphi}$ and $r_{\Phi}=\rho \sin{\varphi}$, the Hamiltonian
reads
\begin{equation}
H_{rel}= \left(- \partial^2_{\rho}  -\frac{1}{\rho}
\partial_{\rho} -\frac{1}{\rho^2} \partial^2_{\varphi}\right) +
\frac{1}{\rho}
\end{equation}
To comply with scattering state boundary conditions, we specify
the orbital collision plane to be formed by the incident
wave-vector $\ks_i$ and the wave-vector $\ks_f$ of the out-going
scattered wave and therefore $\hat{\pc}=\ks_i \times \ks_f$
specifies the orientation of the orbital collision plane, which is
otherwise arbitrary. We impose planar two-dimensional boundary
condition for scattering states as
\begin{equation}
\psi^{(+)}(\rs) \approx \psi^{{\rm inc}}_{\ks_i}(\rs) +
f_S(k,\varphi) \frac{ {\cal F}^{(+)}(k \rho) }{ \sqrt{\rho} },
\quad \rho \rightarrow \infty
\end{equation}
where $\psi^{{\rm inc}}$ is the incident Bloch wave, which is
superimposed on out-going scattered Bloch wave ${\cal F}^{(+)}$ of
amplitude $f(\varphi)$. The planar Bloch wave-functions of
electronic states exhibit partial wave-expansion as
\begin{equation}
\psi_{S,\hat{\pc}}(\rho,\varphi) = e^{i S \varphi} \sum_n e^{2 i n
\varphi} \psi^{(+)}_{2n+S}(k \rho).
\end{equation}
Similarly, the scattering amplitude exhibits Bloch representation
\begin{equation}
f_S(k,\varphi) = e^{i S \varphi} \sum_n e^{2 i n \varphi}
f_{2n+S}(k),
\end{equation}
and satisfies
\begin{equation}
f_S(k,\varphi) = (-1)^S f_S(k,\varphi+\pi),
\end{equation}
i.e. it is symmetric for scattering in a spin-singlet and
anti-symmetric otherwise. Instead for the physical wave-function
$\psi^{(+)}(k \rho)$, we solve this equation for the regular
wave-function $R(k \rho)$, which is subject to the boundary
condition that
\begin{equation}
\lim_{\rho \rightarrow 0} \rho^{-|\Lambda|} R_{|\Lambda|}(\rho)=1
. \label{regular}
\end{equation}
and satisfies the equation
\begin{equation}
\frac{d^2}{d \rho^2} R_{\Lambda}(\rho) + \frac{1}{\rho} \frac{d}{d
\rho} R_{\Lambda}(\rho) +
\left(k^2-\frac{1}{\rho}-\frac{\Lambda^2}{\rho^2} \right)
R_{\Lambda}(\rho)=0, \label{radial}
\end{equation}
where $\Lambda=S~{{\rm mod 2\hbar}}$ is the helicity, $k =
\sqrt{2(\lambda-P^2/8)}$ and the solution depends only on
$|\Lambda|$, and we further take $\Lambda \ge 0$, which is
equivalent to take $n \ge 0$ and consider right-handed electronic
states. Making the substitution $R_{\Lambda} = u_{\Lambda} /
\sqrt{\rho}$ in Eq.(\ref{radial}), we obtain the Whittaker's
equation
\begin{equation}
\frac{d^2}{d z^2} u_{\Lambda}(z) + \left(-\frac{1}{4} +
\frac{\eta}{z} + \frac{1/4-\Lambda^2}{z^2} \right)
u_{\Lambda}(z)=0 ,
\end{equation}
where $z=-2 i k \rho$ and $\eta=-i/2k$. The general solution of
the Whittaker's equation is
\begin{equation}
u_{\Lambda}(z) = A_{\Lambda} W_{\eta,\Lambda}(z) + B_{\Lambda}
W_{-\eta,\Lambda}(-z),
\end{equation}
where $A_{\Lambda}$ and $B_{\Lambda}$ are integration constants
and $W_{\pm \eta, \Lambda}(\pm z)$ are the two linearly
independent Whittaker's functions of second kind. The unknown
integration constants can be obtained from the boundary condition
for the regular solution Eq.(\ref{regular}). The asymptotic of the
Whittaker's functions \cite{spec}, when $|z| \rightarrow \infty$,
\begin{equation}
W_{\eta, \Lambda}(z) \rightarrow z^{\eta} e^{-z /2 },
\end{equation}
gives the asymptotic of the scattering-state wave-function
\begin{eqnarray}
& & u_{\Lambda}(z) \approx e^{\pi / 2 k }
\left[A_{\Lambda} e^{i ( k \rho - \log{2 k \rho} / k )} + \right. \nonumber \\
& & + \left. B_{\Lambda} e^{-i (k \rho - \log{2 k \rho} / k)}
\right], \quad \rho \rightarrow \infty ,
\end{eqnarray}
as linear combination of irregular Jost solutions, specified by
the boundary conditions
\begin{equation}
\lim_{\rho \rightarrow \infty} e^{\mp i [k \rho - \log{2 k \rho} /
k ] } {\cal F}^{(\pm)}(k \rho)= 1,
\end{equation}
and describe polar Coulomb waves outgoing from [with sign $(+)$]
or incoming [with $(-)$ sign] towards the origin $\rho=0$. The
Jost solutions and Whittaker's functions are identical up to
multiplicative $\Lambda$-independent constant, more specifically
the relation is given by
\begin{equation}
{\cal F}^{(\pm)}(k \rho) = e^{\pi/2 k} W_{\pm \eta, \Lambda}(\pm
z),
\end{equation}
and these functions can be identified by $\Lambda$-independent
re-definition of integration constants
\begin{equation}
A \rightarrow A e^{\pi / 2 k }= {\rm f}^{(-)}(k), \quad B
\rightarrow B e^{\pi / 2k }=-{\rm f}^{(+)}(k),
\end{equation}
The wave-function can be re-written as linear combination of the
two Jost solutions
\begin{equation}
u(\rho)=\frac{1}{w(k)} [ {\rm f}^{(-)}(k) {\cal F}^{(+)}(k \rho)-
{\rm f}^{(+)}(k) {\cal F}^{(-)}(k \rho) ],
\end{equation}
where $w(k)=W[{\cal F}^{(-)},{\cal F}^{(+)}]=2 i k$ is the
Wronskian of the two Jost solutions, i.e. $W[f,g]=fg'-f'g$ and
prime denotes radial derivative. The integration constants ${\rm
f}^{(\pm)}$ are the Jost functions, which are given by the
Wronskians
\begin{equation}
{\rm f}^{(\pm)}(k) = W[{\cal F}^{(\pm)},u],
\end{equation}
which we evaluate at the origin $\rho=0$. By using the asymptotic
of the irregular solutions near the origin
\begin{equation}
\lim_{|z| \rightarrow 0} z^{\Lambda-1/2} W_{\eta, \Lambda}(z) =
\frac{\Gamma(2\Lambda)}{\Gamma(\Lambda + i/2k + 1/2)}
\end{equation}
and evaluating Wronskians with the help of the boundary condition
in Eq.(\ref{regular}), we obtain Jost functions as
\begin{equation}
{\rm f}^{(\pm)}(k) = e^{\pi/2k} (\mp 2i k)^{1/2-\Lambda}
\frac{\Gamma(2\Lambda+1)}{\Gamma(\Lambda \pm i/2k + 1/2)},
\end{equation}
where $\Gamma(z)$ is the Euler's gamma function. Then the
asymptotic scattering-state wave-function $\rho \rightarrow \infty
$ is given by
\begin{eqnarray}
& & u_\Lambda(\rho) \approx  \frac{ e^{\pi/2k} (2
k)^{-1/2-\Lambda}
\Gamma{(2\Lambda+1)}}{|\Gamma(\Lambda+1/2+i/2 k) |} e^{i \delta_{\Lambda}} \times \nonumber \\
& & \times \sin \left(k \rho - \frac{1}{k} \log{2 k \rho}
-\frac{\pi \Lambda}{2} - \frac{\pi}{4}+  \delta_{\Lambda} \right)
, \label{sol}
\end{eqnarray}
where $\delta_{\Lambda}$ are elastic scattering phase shifts
\begin{equation}
\delta_{\Lambda}(k)={\rm arg}\Gamma(\Lambda+1/2+ i/2k),
\end{equation}
relative to the asymptotic of the non-interacting Bessel's
function $J_{\Lambda}(k \rho)$,
\begin{equation}
J_{\Lambda}(z) \approx \sqrt{ \frac{2}{\pi z} } \cos\left( z
-\frac{\pi \Lambda}{2} - \frac{\pi}{4} \right). \label{bessel}
\end{equation}
The Coulombic $S$-matrix is given by the factor of the Jost
functions
\begin{equation}
S_{\Lambda}(k) = \frac{ {\rm f}^{(-)}(k)}{ {\rm f}^{(+)}(k)} =
e^{2 i \delta_{\Lambda}(k)}.
\end{equation}
The physical wave-function differs from the regular wave-function
be a normalization constant determined from the Jost function,
i.e.
\begin{equation}
\psi^{(+)}_{\Lambda}(k \rho) = N_{\Lambda}(k) \frac{u_{\Lambda}(k
\rho)}{\sqrt{2 k \rho}},
\end{equation}
and therefore
\begin{equation}
N_{\Lambda}(k) = e^{-\pi/2k} (2 k)^{\Lambda} \frac{
\Gamma(\Lambda+1/2+i/2k ) } {(2\Lambda) ! }.
\end{equation}
As a result, we obtain the partial-wave scattering amplitudes
$f_{\Lambda}$ as
\begin{equation}
f_{\Lambda}(k) = \frac{e^{-i \pi/4}}{\sqrt{2 \pi k} } [e^{2i
\delta_{\Lambda}(k)} -1].
\end{equation}
We next evaluate a differential cross-section for scattering in a
given line segment in the collision plane as
\begin{equation}
d P_{\varphi} = |f(\varphi)|^2  d \varphi.
\end{equation}
To evaluate  outgoing scattered probability flux through a solid
angle $d \Omega$, we vary the wave-vector $\ks_f$, such that the
unit vector $\ks_i \times \ks_f$ normal to the scattering plane
rotates on angle $\alpha$ about the axis of incidence $\ks_i$, and
a generated linear flux is
\begin{equation}
d P_{\alpha} = |f(\varphi)|^2 (\sin{\varphi} d \alpha).
\end{equation}
Therefore elastic scattering cross-section is given by
\begin{equation}
d \sigma = d P_{\varphi} d P_{\alpha}= |f(\varphi)|^4 d \Omega,
\end{equation}
where $d \Omega= \sin{\varphi} d \varphi d \alpha$ is the solid
angle of observation in the space-fixed reference frame. The
differential cross-section is given by
\begin{equation}
\frac{d \sigma}{d \hat{\ks}_f} =\frac{d \sigma(\hat{\ks}_f
\leftarrow \hat{\ks}_i)}{d \Omega} =|f(k, \varphi)|^4,
\end{equation}
and exhibits characteristic dependence on the fourth power of the
amplitude.

\subsection{Quasi-classical approximation}
In quasi-classical approximation, the two-electron wave-function
is a phase-factor
\begin{equation}
\psi = e^{i S},
\end{equation}
expanding the phase $S=S_0+\hbar S_1 + \ldots$, to zero'th order
in the Planck's constant, we obtain the equation of motion for the
active electron as
\begin{equation}
\frac{1}{2} [\nabla_{\rs_1} S_0(\rs_1,\rs_2)]^2 + g
|\rs_1-\rs_2|^{-1} = \lambda,
\end{equation}
together with the constraint for particle identity
\begin{equation}
[\nabla_{\rs_1} S_0(\rs_1,\rs_2)]^2 =  [\nabla_{\rs_2}
S_0(\rs_1,\rs_2)]^2.
\end{equation}
The quasi-classical Coulmbic action has the form
\begin{equation}
S_0(\rs, \rc) = \pc \cdot \rc + \sigma_{\pc}(\rs_{\bot}),
\end{equation}
where the action for the relative motion satisfies the equation
\begin{equation}
[\nabla_{\bot} \sigma_{\pc}]^2 + \frac{1}{r_{\bot}} = k^2,
\end{equation}
where $1/r_{\bot}$ is the screened planar Coulomb potential.
Re-introducing plane polar coordinates $\rs_{\bot}=(\rho,\varphi)$
to describe a collision, the quasi-classical equation reduces
exactly to the Hamilton-Jacobi equation of the planar Kepler
problem
\begin{equation}
[ \partial_{\rho} \sigma]^2 + \frac{1}{\rho^2}[\partial_{\varphi}
\sigma]^2 + \frac{1}{\rho} = k^2. \label{Jacobi}
\end{equation}
and has solutions
\begin{equation}
\sigma = \pm \Lambda \varphi + \sigma_{\Lambda}(\rho),
\end{equation}
where $\Lambda$ labels the helicity. The quasi-classical
approximation holds if $\Lambda \gg 1$, and therefore we will
neglect effects of quantization of $\Lambda$. The planar Kepler
problem exhibits dynamical symmetry (e.g. \cite{LRL1,LRL2}), due
to the conservation of the planar Laplace-Runge-Lenz vector, which
is given by (we use the classical expression, which does not
involve hermitian symmetrization)
\begin{equation}
\ac=\ps \times \ls + \hat{\rs},
\end{equation}
where $\ls=\rs \times \ps$ is the relative angular momentum,
$\ps=\nabla_{\rs} \sigma(\rs_{\bot})$ is the relative
quasi-classical momentum and $\ac \cdot \ls =0$. The classical
trajectories of relative motion of the two-electrons can be
obtained from
\begin{equation}
\rs \cdot \ac= \rho A \cos{\varphi} = \Lambda^2+\rho
\end{equation}
which leads to the conical section equation, specifying unbound
hyperbolic Kepler orbits
\begin{equation}
\frac{p}{\rho}= -1 + e \cos{\varphi},
\end{equation}
with parameter $p=\Lambda^2$ and eccentricity $e=A=\sqrt{1+k^2
\Lambda^2}>1$, and the point of closest approach on the trajectory
($\varphi=0$) is $\rho_{min}=p/(e-1)$. The cross-section for
elastic scattering can be derived from the conservation of the
Runge-Lenz vector. In a reference frame, where the center-of-mass
motion is at rest $\pc={\bf 0}$, we choose the direction of
incidence $\hat{\ks}_i$  be the negative half of the $x$-axis of
the laboratory frame, the asymptotic momentum of relative motion
is $\ks_i= k \hat{e}_x$, the angular momentum of relative motion
corresponding to this choice is $\Lambda= x p_y-y p_x =-b k$,
where $b$ is the impact parameter, and $\hat{e}_y$ is a unit
vector in the plane of the orbit. The Laplace-Runge-Lenz vector
prior to the collision is given by
\begin{equation}
\ac_{in}=  -\hat{e}_x + b k^2 \hat{e}_y,
\end{equation}
and similarly after the collision
\begin{equation}
\ac_{out}= \hat{e}_{out} + b k^2 \hat{\ns}_{out},
\end{equation}
where $\hat{e}_{out}=\hat{\rs}_{\bot}$ is a unit vector specifying
the outgoing direction of the scattered particles,
$\hat{\ns}_{out}=\hat{\pc} \times \hat{e}_{out}$ is a unit-vector
in the collision plane, and $\hat{\pc}$ is the unit vector normal
to the plane of the orbit. By projecting the conserved Runge-Lenz
vector onto the direction of incidence, i.e.
\begin{equation}
\hat{e}_x \cdot \ac_{in} = \hat{e}_x \cdot \ac_{out},
\end{equation}
we obtain that
\begin{equation}
-1 =  \cos{\chi} -b k^2 \sin{\chi}, \label{LRL}
\end{equation}
where $\chi$ is a rotation angle, $\cos{\chi}=\hat{e}_x \cdot
\hat{e}_{out}$. From Eq.(\ref{LRL}) we obtain the relation
$b=b(\chi)$ between the deflection angle and the impact parameter
as
\begin{equation}
b =\frac{1}{k^2 \tan{\chi/2}}. \label{impct}
\end{equation}
Differentiating with respect to $\chi$, we obtain
\begin{equation}
\left| \frac{d b}{d \chi} \right| = \frac{1}{2 k^2 \sin^2{\chi/2}}
\label{jacob}.
\end{equation}
The classical differential scattering cross-section is given by $d
\sigma = 2 \pi b d b$, and by using that $d\Omega=2 \pi \sin\chi d
\chi$, together with Eq.(\ref{impct}) and Eq.(\ref{jacob}) we
obtain the Rutherford formula for the cross-section
\begin{equation}
\frac{d \sigma}{d \Omega} = \frac{1}{4 k^4 \sin^{4}{\chi/2}}.
\end{equation}
Taking into account reflection symmetry $\chi \rightarrow
\pi-\chi$, i.e. that we can not distinguish between forward and
backward scattering when the particles are identical, we obtain
that
\begin{equation}
\frac{d \sigma(\pi-\chi)}{d \Omega} = \frac{1}{4 k^4
\cos^{4}{\chi/2}}
\end{equation}
The total cross-section for elastic scattering is obtained by the
sum of the two contributions
\begin{equation}
\frac{d \sigma}{d \Omega} = \frac{1}{4 k^4} \left(
\frac{1}{\sin^{4}{\chi/2}}+\frac{1}{\cos^{4}{\chi/2}}\right)
\end{equation}
The cross-section for scattering at small and large angles is
highly divergent. The cross-section can be defined only when the
center-of-mass motion is at rest. The particles are identical and
have equal kinetic energies $p_1^2 =p_2^2$ due to the statistics,
and there is no reference frame where only one of the particles is
at rest. This also means that the problem does not exhibit
spherical rotation symmetry, instead it exhibits cylindrical
rotation symmetry.

\section{Conclusion}
We show that in the particular case of systems with two particles,
that the constraints of particle identity entail reduction in the
number of internal degrees-of-freedom from six to five. The effect
of redundancy in the description of orbital motion in the
two-particle gauge system is found to be in correspondence with
the multiplicative phase-factor $(-1)^S$, where $S=\{0,1\}$ is the
total spin.

\section{Acknowledgment}

\end{document}